\documentstyle[10pt,emulateapj]{article}
\lefthead{FURTON, LAIHO \& WITT}
\righthead{INTERSTELLAR HYDROGENATED AMORPHOUS CARBON}
\newcommand{\apl}{Appl. Phys. Lett.}
\newcommand{\jvst}{J. Vac. Sci. Technol.}
\newcommand{\jap}{J. Appl. Phys.}
\newcommand{\etal}{{\it et al.}}
\newcommand{\icm}{cm^{-1}}
\newcommand{\chcm}{\sim\!\!2950\;\icm}
\newcommand{\EE}[1]{\times10^{#1}}
\newcommand{\Plam}{\Psi (\lambda)}
\newcommand{\Dlam}{\Delta (\lambda)}
\newcommand{\nlam}{n (\lambda)}
\newcommand{\klam}{k (\lambda)}
\begin{document}
\title{The Amount of Interstellar Carbon Locked in Solid Hydrogenated
Amorphous Carbon}
\author{Douglas G. Furton, John W. Laiho}
\affil{Rhode Island College}
\authoraddr{Department of Physical Sciences, Providence, RI 02908}
\and
\author{Adolf N. Witt}
\affil{University of Toledo}
\authoraddr{Department of Physics \& Astronomy, Toledo, OH 43606}

\begin{abstract}

Some form of hydrogenated amorphous carbon (HAC) is widely regarded as
a likely component of dust in the diffuse interstellar medium of the
Galaxy.  The most direct observational evidence is the $\chcm$ C-H
stretch feature seen in absorption along heavily attenuated lines of
sight. We have used the measured properties of one well-characterized
HAC sample, together with well-established correlations between the
structural, optical, and infrared (IR) characteristics of HAC
materials in general, to estimate the amount of carbon relative to
hydrogen that must be bound in this particular solid in interstellar
space. The solution to this problem remains uncertain because of the
strong dependence of the $\chcm$ mass-absorption coefficient upon the
H/C ratio in HAC materials. We estimate that about $80\;ppM$ carbon
relative to hydrogen is locked up in interstellar HAC grains with
H/C$\;\simeq 0.5$. A H/C ratio lower than this value would require
higher carbon depletions and result in a $\chcm$ absorption profile
inconsistent with observations. Substantially higher H/C ratios, which
would imply carbon depletions as low as $20\;ppM$, would lead
to materials exhibiting efficient blue-green photoluminescence, which
is not observed. Combined with observations of the gas-phase
interstellar abundance of carbon and the revised interstellar
reference abundances of refractory elements, this result implies that
the bulk of solid carbon in interstellar space is in the form of HAC.
In addition to a careful analysis of the IR properties of HAC, we
present a summary of the UV/visible optical properties, including the
complex index of refraction, measured for one particularly
well-characterized HAC sample that is shown to have an IR absorption
spectrum which is {\em quantitatively} consistent with the latest IR
observations of the diffuse interstellar medium.
\end{abstract}


\section{Introduction}

It is now well established on the basis of astronomical and
laboratory-based infrared (IR) spectroscopy that some form of solid,
hydrogenated carbonaceous material is present in the diffuse
interstellar medium (ISM).  The absorption feature at $\chcm$
characteristic of the C-H stretching mode in hydrogenated carbonaceous
materials is clearly and consistently revealed by observational
studies of dusty lines of sight, from early near-IR spectrophotometric
observations of the Galactic center by Soifer, Russell, \& Merrill
(1976)\markcite{srm76} to the most recent observations of
Cygnus~OB2~\#12, the prototypical diffuse ISM line of sight, by
Whittet \etal~(1997)\markcite{w97} using the Infrared Space
Observatory (ISO).  Other relevant observational works include: the
observation of Cygnus~OB2~\#12 by Adamson, Whittet, \& Duley
(1990)\markcite{awd90}, the near-IR spectral surveys of Galactic
center sources and other diffuse ISM probes by Sandford
\etal~(1991)\markcite{s91} and Pendleton \etal~(1994)\markcite{p94},
and the near-IR spectroscopy of the proto-planetary nebula CRL~618 by
Chiar \etal~(1998)\markcite{c98}.  Without considering observational
and laboratory work outside the $3500\!-\!2500\;\icm$ spectral region,
the spectral profile of the $\chcm$ C-H stretching feature itself is
not consistent with any simple hydrocarbon molecule -- including
polycyclic aromatic hydrocarbons (PAHs) -- or mixtures thereof (e.g. Bellamy
1975)\markcite{b75}.  It is the signature of some form of cold
(but not frozen) hydrogenated carbonaceous material which we will call
HAC.

HACs comprise a wide range of materials composed only of carbon and
hydrogen, but which vary in $sp^1\!\!:\!\!sp^2\!\!:\!\!sp^3$
hybridization ratios and H/C ratio.  There is substantial empirical
and experimental evidence to indicate that for C-H networks the carbon
hybridization ratios, which determine the structural and optical
properties of the material, and the H/C ratio are related; they do not
vary independently. We point to the reviews of HAC by Robertson
(1986)\markcite{r86}; the ``random covalent network'' model by Angus
\& Jansen (1988)\markcite{aj88}; the ``defected graphite'' model by
Tamor \& Wu (1990)\markcite{tw90}; and the experimental
characterizations of various HACs by Dischler, Bubenzer \&
Koidl~(1983)\markcite{dbk83}; McKenzie \etal~(1983)\markcite{m83};
Tsai \& Bogy~(1987)\markcite{tb87}; Gonzalez-Hernandez, Chao \&
Pawlik~(1988)\markcite{gcp88}; and Tamor, Wu
\etal~(1989)\markcite{tw89}.  It is clear from these works that for
HAC in general, regardless of production method: 1)~the $sp^3/sp^2$
ratio, optical gap and the real part of the index of refraction all
increase with increasing H/C ratio; 2)~the density and the imaginary
part of the index of refraction both decrease with increasing H/C
ratio; 3)~HACs seldom contain a significant concentration of $sp^1$
hybridized carbon; 4)~the maximum value of H/C is near 1.5; and
5)~HACs photoluminesce with an efficiency and energy of peak emission
that depend on the H/C ratio.  In total, these works demonstrate that
amorphous, carbonaceous solids in general have predictable,
well-determined optical and physical properties, despite having
fundamental parameters like the H/C ratio and the $sp^3/sp^2$ ratio
which are wide ranging.  Thus it is possible and indeed practical to
investigate and draw conclusions about the interstellar HACs from HAC
analogs which are produced by methods and under conditions far
different than the interstellar case.  So, while the astrophysical
production method of HAC may be the photolysis of organic ices as
proposed by Bernstein \etal~(1995)\markcite{b95} or Greenberg
\etal~(1995)\markcite{g95} (among others), or may be by direct
deposition onto silicate cores as suggested by Jones, Duley, \&
Williams (1990)\markcite{jdw90}, results of experiments on HAC
materials produced by plasma-enhanced chemical-vapor deposition
(PECVD), arc-evaporation, or laser ablation, for example, are directly
applicable to the astrophysical problem under consideration.

It is imperative to determine more precisely the nature of
interstellar HAC in the light of recent reconsiderations of the
abundance of solid- and gas-phase carbon in the diffuse ISM.  It now
seems likely that the cosmic abundance of carbon (and other heavy
elements) is perhaps as low as $0.5\!-\!0.75$ of the solar abundance,
which had long been thought to be representative of abundances in the
ISM.  Snow \& Witt (1995)\markcite{sw95} review the existing
literature on the carbon abundance in the sun, in a number of recently
formed B stars, and in F and G stars similar to the sun, and conclude
that the abundance of carbon relative to hydrogen in the ISM is
$225\!\pm\!50\;ppM$.  In addition, the amount of carbon in the gas
phase has recently been redetermined more precisely by Cardelli
\etal~(1996)\markcite{c96} through observations with the Goddard
High-Resolution Spectrograph of the weak C II] $\lambda2325\;{\rm
\AA}$ absorption line along lines of sight with greatly different
ratios of atomic to molecular hydrogen.  They derive a gas-phase
carbon abundance relative to hydrogen of $140\!\pm\!50\;ppM$.  Thus it
appears that the amount of carbon available to form solid material in
the ISM -- a parameter that is crucial to all dust-models of
interstellar extinction (e.g., Mathis 1996\markcite{m96}; Li \&
Greenberg 1997\markcite{lg97}) -- is likely to be only about $80\;ppM$
of hydrogen, to within an uncertainty of about a factor of two.

Thus, the first goal of this current work is to present a {\em
rigorous} determination of the portion of the solid-phase of carbon in
the diffuse ISM that must be locked in the form of HAC to give rise to
the observed absorption in the interstellar $\chcm$ C-H stretching
feature.

Further, it is important to understand the role interstellar HAC
grains play in the extinction of starlight, given the presence of HAC
in the general diffuse ISM. In order to proceed, the complex index of
refraction of HACs as a function of energy in the ultraviolet (UV) and
visible regions of the electromagnetic spectrum must be known, in
addition to the amount of this material that is present. Assumptions
about the structural nature of the grains (i.e., core/mantle or
uniform composition) and the size/shape distributions for the grains
must also be made, but laboratory investigation is of little help
here.  Tabulated optical constants of materials presented as analogs
to the diffuse ISM carbonaceous material, ranging from ``organic
refractory'' materials to HACs and other forms of amorphous carbon,
are plentiful in the literature and are used as needed in models of
interstellar extinction.  For example, see Smith (1984)\markcite{s84};
Duley (1984)\markcite{d84}; Bussoletti \etal~(1987)\markcite{b87};
Alterovitz \etal~(1991)\markcite{a91}; Jenniskens
(1993)\markcite{j93}; Colangeli (1995)\markcite{c95}; and Schnaiter
\etal~(1998\markcite{s98}, 1999\markcite{s99}).

The list of ``optical constants'' works
just cited, it is important to note, does not completely overlap with 
the list of ``IR spectroscopy papers'' cited in the previous 
paragraphs.  So, while there has been a great deal of work in the 
areas of IR and UV/visible spectroscopies of HAC materials, the works 
are in large measure disjoint.  The very recent works of Jaeger 
\etal~(1998)\markcite{j98} and Schnaiter \etal~(1998\markcite{s98}, 
1999\markcite{s99}) deserve special mention as they begin to consider 
the problems of grain composition, structure, and optical properties 
collectively.

A second goal of this current work, then, is to present the complex
index of refraction as a function of energy in the UV/visible for a
HAC material with an IR absorption spectrum that is {\em
quantitatively} consistent with the latest IR observations of dust in
the diffuse ISM, particularly the ISO spectrum of Cyg.~OB2~\#12 by
Whittet \etal~(1997)\markcite{w97}, a data set that was until now not
available in the literature.

In addition to absorbing and scattering starlight, it has been
suggested that HAC grains are the carrier of the broad, red emission
band known as extended red emission (ERE) (e.g., Duley
1985\markcite{d85}; Witt \& Schild 1988\markcite{ws88}; Witt
1994\markcite{w94}). ERE has been detected in a variety of dusty
astrophysical environments, including: reflection nebulae, planetary
nebulae, dark nebulae, Galactic cirrus clouds, an H II region, the
halo of M82, and a nova.  Most recently, Gordon, Witt, \& Friedmann
(1998)\markcite{gwf98} have shown that ERE is also a phenomenon
associated with general diffuse interstellar medium dust (see this
work for references to the numerous original ERE detections).  They
also conclude that the quantum efficiency of the ERE process is in
excess of 10\%.  Witt, Gordon, \& Furton (1998)\markcite{wgf98} argue
that this extremely high efficiency, among other things, forces a
reconsideration of the notion that ERE is due to interstellar HAC
grains.  This same argument has also been made by Ledoux
\etal~(1998)\markcite{l98}.  At the very least, this finding demands a
careful measurement of the photoluminescence (PL) efficiency of HAC
materials that are otherwise viable analogs to the interstellar
carbonaceous material.

The experimental aspects of this work concerning the PECVD of and
subsequent characterization by UV/visible/IR absorption
spectroscopies, optical PL spectroscopy, and UV/visible spectroscopic
ellipsometry are described in \S2.  A rigorous determination of the
amount of carbon locked in interstellar HAC grains is presented in
\S3.1. The characteristics that any viable interstellar HAC analog
must posess, in light of the analysis presented in \S3.1, are
discussed in \S3.2. The optical role HAC grains are likely to play in
the extinction of UV/visible starlight, and in the dust luminescence
processes known as ERE in the diffuse ISM is considered in \S3.3. All
of this is followed by a brief conclusion in \S4.

\section{Experimental Considerations, Analysis and Results}

In our laboratories at the University of Toledo and Rhode Island
College, we have been involved in depositing and characterizing thin
films of HAC and related materials for almost a decade.  We use both
RF- and DC-based PECVD systems to produce thin, solid films from
mixtures of gas-phase precursors.  We rely on UV/visible/IR absorption
spectroscopies, visible and near-IR PL spectroscopy, among other
techniques such as spectroscopic ellipsometry and electron microscopy,
for analysis and characterization of these thin-film materials. In
this work, we present a thorough characterization of a single HAC
sample, from among the many samples we have produced and analyzed, as
a viable analog to the interstellar hydrogenated, carbonaceous
material.  The deposition method and conditions, along with a summary
of the derived optical and physical properties are presented in \S2.1.
The measurement and analysis that were completed to derive the
physical and optical properties of this sample, including its
electronic band-gap, density, H/C ratio, $sp^3/sp^2$ ratio, PL
efficiency, complex index of refraction, and IR
($4000\!-\!1000\;\icm$) mass-absorption coefficient are presented in
\S\S2.2-2.5.

\subsection{HAC Sample Preparation and Summary of Derived Physical and
Optical Properties}

This interstellar HAC analog was produced using a DC-based PECVD
system of a somewhat unique design.  Although DC PECVD is
electronically simpler than RF PECVD, it does not lend itself as well
to depositing materials (either conducting or insulating) onto
insulating substrates, which are most commonly used (i.e., fused
silica and salt).  The system we have designed minimizes this problem.
It consists of a vacuum chamber pumped by a roughing pump and a
diffusion pump, with a four-channel precursor gas mixing manifold.
Electronically, the chamber itself is grounded and the sample is
deposited onto the substrate placed on a support about $5\;mm$ beneath
a $3.0\;cm$ diameter nickel-chromium screen which serves as the
cathode.  All of this is surrounded by a chimney-like glass tube about
$7\;cm$ in diameter, open at the bottom and very near the top of the
chamber, in order to confine the plasma and to minimize possible
contamination of the deposited samples by contaminants on the metal
walls of the vacuum chamber.  During the deposition process, the
conducting screen is pulled to a negative DC potential which is
variable between $0\!-\!2000\;V$ by a high-voltage power supply and a
current-limiting resistor.  The screen maintains the plasma, which
causes deposition onto the substrate just below as ionized molecular
fragments of the precursor gas mixture ``overshoot'' it.  The screen
itself is coated during the deposition process as well. For this
reason, a clean screen is used for each deposition and, as a drawback,
the deposition process is self-limiting because the current through
the plasma into the screen decreases as the screen becomes coated by
the insulating material, eventually becoming too low to support the
plasma.  HAC films can be deposited with thickness up to about
$200\;nm$ in a single deposition lasting about $30\;min$; thicker
films can be deposited by interrupting the process to replace the
cathode screen.

In general, samples deposited under identical conditions have 
indentical physical and optical properties.  The materials that result
are thin, solid, homogeneous films; they are not porous or particulate
in nature.

The deposition conditions for the HAC sample presented in this work
are as follows:  the precursor gas was 99.99\% pure methane, at a
pressure of $200\;mTorr$ with a flow-rate of $5.0\;sccm$; the
electric potential was $-1000\;VDC$ which established a current of
$3.0\;mA$ initially.  The total deposition time was $60\;min$,
consisting of two sequential $30\;min$ depositions between which the
cathode screen was changed.  The several films produced in this way
turned out to be about $300\;nm$ thick.  A sodium chloride substrate
was used for subsequent IR spectroscopic analysis of the sample; a
fused silica substrate was used for subsequent UV/visible analyses.
There was no indication that the HAC material deposited differently
onto the salt substrate than onto the fused silica substrate.

The UV/visible absorption spectrum of this HAC sample was recorded
over the wavelength range $900\!-\!190\;nm$ with a Perkin-Elmer 552
dual beam spectrophotometer; the absorbance spectrum and a
corresponding ``Tauc'' plot, as described in \S2.2, are shown in
Figure~\ref{f1}.  The optical PL spectrum covering the
$500\!-\!950\;nm$ wavelength range was recorded with an Ocean Optics
S2000 fiber-fed CCD spectrometer using the $488\;nm$ line of an
argon-ion laser as the source of excitation.  The PL spectrum shown in
Figure~\ref{f2} was reduced to quantum efficiency as described in
\S2.3. The sample was analyzed by spectroscopic ellipsometry over the
wavelength range $300\!-\!1000\;nm$ with a J.A. Woollam Company
SASE-1.1 ellipsometer.  This analysis, which is discussed in much more
detail in \S2.4, was completed with the help of Dr. Margaret Tuma of
the NASA Lewis Research Center; we are grateful for her assistance.
The indices of refraction ($n$ and $k$) derived from this analysis are
shown in Figure~\ref{f3}.  Finally, the IR absorption of this sample
was recorded over the frequency range $4000\!-\!1000\;\icm$ with a
Nicolet 510P FT-IR spectrometer (the low-frequency limit is imposed by
the salt substrate); this spectrum, reduced to mass-extinction
coefficient as described in \S2.5 is shown in Figure~\ref{f4}.

A summary of the physical and optical properties of this HAC sample,
derived from the measurements indicated in the previous paragraph is
presented in Table~\ref{t1}.  The details of each analysis are presented
in the following subsections.

\begin{deluxetable}{lll}
\tablecaption{Summary of Interstellar HAC Analog Physical and
Optical Properties \label{t1}}
\tablehead{\colhead{Property} & \colhead{Value} & \colhead{Method}}
\startdata
Density     & $1.5 \pm 0.1 \;g/cm^3$ & from sample mass and thickness \nl
Band-gap    & $1.9 \pm 0.1 \;eV$     & Tauc (1973)\markcite{t73} method \nl
H/C ratio   & $0.5 \pm 0.1$          & Jacob \& Unger (1996)\markcite{ju96} \nl
$sp^3/sp^2$ & $0.5 \pm 0.1$          & from H/C and Fig.~4 Tamor \& Wu (1990)
\markcite{tw90}\nl
$\kappa^{\prime}(2950)$ & $1.4 \pm 0.1 \EE{3}\;cm^2/g$ & IR spectrum, density, and
thickness \nl
PL QE & $0.05 \pm 0.02$ & PL spectrum and UV/vis absorption \nl
\enddata
\tablecomments{The thickness of the sample for which these properties
are derived is $300\;nm$; $\kappa^{\prime}(2950)$ is the maximum
mass-extinction in the $\chcm$ absorption feature as shown in
Figure~\ref{f4}.}
\end{deluxetable}

\subsection{HAC Sample Band-gap and Physical Properties}

\begin{figure*}
\plotone{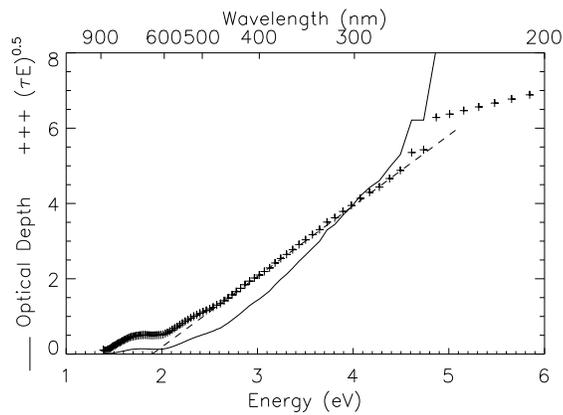}
\caption{Optical depth spectrum and ``Tauc plot'' (see text) for the
interstellar HAC analog.  The $x$-intercept of the dashed line, the
electronic band-gap, is $1.9\;eV$. \label{f1}}
\end{figure*}

A number of laboratory studies have demonstrated the electronic
band-gap of HAC to be a useful parameter related to many of its
optical and physical properties 
(e.g., Robertson 1986\markcite{r86}; Angus \& Jansen 
1988\markcite{aj88}; Tamor, Haire \etal~1989\markcite{th89}; Tamor \& 
Wu 1990\markcite{tw90}; and Witt, Ryutov, \& Furton 
1997\markcite{wrf97}).  The electronic band-gap of any amorphous 
semiconductor is conveniently characterized via the method outlined by 
Tauc (1973)\markcite{t73}.  This method prescribes that the band-gap 
is equal to the $x$-intercept of a plot of $(\alpha E)^{\frac{1}{2}}$ 
versus $E$ ($\alpha$ is the absorption coefficient, $E$ is energy), 
provided that the conduction and valence band edges are parabolic in 
energy, a condition that has proven to be more or less satisfied for 
most HACs.  Figure~\ref{f1} shows a plot of the UV/visible optical 
depth spectrum for the HAC sample, along with a plot of $(\tau 
E)^{\frac{1}{2}}$ derived from the optical depth spectrum according to 
$\tau = \alpha t = -\ln (\frac{I}{I_{o}})$.  The $x$-intercept of the 
$(\tau E)^{\frac{1}{2}}$ plot is $E_g = 1.9\;eV$.  Note that plots of 
$(\tau E)^{\frac{1}{2}}$ and $(\alpha E)^{\frac{1}{2}}$ versus $E$ 
differ only in slope, not in $x$-intercept.\footnote{Note also that 
$\alpha t = -\ln (\frac{I}{I_{o}})$ is only an approximate relation; 
$\alpha$ really needs to be determined more carefully.  The use of 
this approximation in the ``Tauc plot'' is justified, however, given 
the assumptions made in this formalism.}

One physical property that is well-parameterized by $E_g$ is 
mass density.  For example, Figure~3 in Tamor, Wu, 
\etal~(1989)\markcite{tw89} reveals an approximately inverse 
relationship between mass density and band-gap.  The density of HAC 
with $E_g = 1.9\;eV$, the case here, is about $1.5\;g/cm^3$.  We also 
verified this by direct measurement; the mass of HAC deposited onto 
the substrate was determined with a microbalance (mass after 
deposition minus mass before deposition), then with the sample 
thickness determined by the ellipsometric analysis described below, 
the HAC sample density of $1.5\;g/cm^3$ was verfied to within the 
precision of the measurements.  The uncertainty in this measurement is 
approximately $0.1\;g/cm^3$.

A review of the literature shows that the $sp^3/sp^2$ carbon 
hybridization ratio, the H/C atom ratio, and the band-gap all are well 
correlated for HACs in general.  Table~I in Kaplan, Jansen, \& 
Machonkin (1985)\markcite{kjm85}; Figure~7 in Angus \& Hayman 
(1988)\markcite{ah88}; and Figure~4 in Tamor \& Wu 
(1990)\markcite{tw90} are representative of this correlation.  From 
these figures we determine that $sp^3/sp^2 = 0.5 \pm 0.1$ and H/C$\;= 
0.5 \pm 0.1$ for this HAC sample with $E_g = 1.9\;eV$.  The estimated 
errors for these quantities stem from the uncertainty in $E_g$, and 
from the precision with which the data in the tables and figures cited 
above can be interpolated.  We also determined the H/C atom ratio 
directly from an analysis of the C-H stretch absorption feature via 
the method prescribed by Jacob \& Unger (1996)\markcite{ju96}; the
detailis of this derivation are desribed in \S2.5.  We find,
consistent with the results above, that H/C$\;\sim 0.5$.

\subsection{HAC Sample Photoluminescence}

It is well known that HACs luminesce upon exposure to blue and near-UV
radiation, with an overall efficiency and energy of peak emission that
depend strongly and directly on the band-gap of the material (e.g.,
Watanabe, Hasegawa, \& Kurata 1982\markcite{whk82}; Silva, 
\etal~1996\markcite{srr96}; Witt, Ryutov, \& Furton 1997\markcite{wrf97}).

\begin{figure*}
\plotone{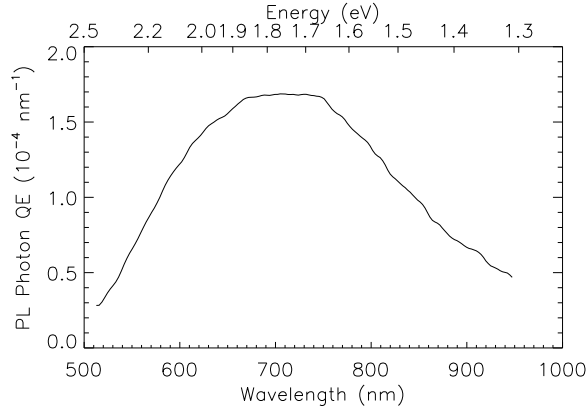}
\caption{The optical photoluminescence photon quantum efficiency (QE)
of the interstellar HAC analog. The integrated PL QE for
this material is $0.05 \pm 0.02$. \label{f2}}
\end{figure*}

The optical PL spectrum for this HAC sample was recorded over the
wavelength range $500\!-\!950\;nm$ with an Ocean Optics S2000
fiber-fed CCD spectrometer.  The excitation source was the $488\;nm$
line of an argon-ion laser.  This fiber/spectrometer combination has
been calibrated to absolute units via a NIST-traceable standard lamp.
Furthermore, the spectrum, which is shown in Figure~\ref{f2}, has been
reduced to approximately represent the PL photon quantum efficiency
(QE) by dividing the observed PL photon count by the number of
absorbed excitation laser photons, as computed from the measured laser
intensity and the measured sample absorbance at the laser wavelength.
This method is approximate because the radiative transport of the
exciting and emitted photons, including scattering and
self-absorption, is not accounted for.  The error introduced by this
simplification is smaller than the other errors associated with the PL
observation and absolute calibration.  Overall, this HAC sample has an
integrated PL QE of $0.05 \pm 0.02$, with a wavelength of peak
emission around $700\;nm$.

\subsection{HAC Sample Index of Refraction}

The complex index of refraction ($n - ik$) of this HAC sample and its
thickness were determined over the wavelength range $300\!-\!1000\;nm$
by spectroscopic ellipsometry.  In this techniqure, the relative
amplitudes and phase-shifts between parallel and perpendicularly
polarized light reflected from the surface of a thin film sample are
measured as functions of wavelength; these functions are denoted
$\Plam$ and $\Dlam$, respectively.  Rather sophisticated modeling is
required, however, to reduce the $\Plam$ and $\Dlam$ data to the
sample thickness and its optical constants $\nlam$ and $\klam$ (Aspnes
1985\markcite{a85}).  Despite this difficulty, spectroscopic
ellipsometry is arguably the best technique to determine index of
refraction for thin-film samples, and one which has not been applied
to HAC materials in great measure.

The ellipsometric analysis of this HAC sample (the measurement of
$\Plam$ and $\Dlam$) was performed with a J.A. Woollam Company
SASE-1.1 spectroscopic ellipsometer by Dr. Margaret Tuma at the NASA
Lewis Research Center.  The reduction of $\Plam$ and $\Dlam$ to
$\nlam$ and $\klam$ (and sample thickness) was completed by one of the
authors (JWL); the method of analysis is described in some detail
below since the results are heavily model dependent.

Equations describing the reflection and transmission of light incident
at some angle on a thin film of an optically isotropic and homogeneous
material with complex index of refraction $n - ik$ deposited on a
substrate are derived from Maxwell's equations.  The amplitude and
phase of the reflected beam, which are analyzed by the ellipsometer,
are different for light polarized parallel to the plane of incidence
than for light polarized perpendicular to the plane of incidence.  The
ratio $\frac{R_\|}{R_\bot}$ defines two functions of wavelength --
$\Plam$ and $\Dlam$ -- according to the relation

\begin{equation}
\frac{R_\|(\lambda)}{R_\bot(\lambda)} = e^{i \Dlam}\; \tan \Plam
\;\;. \label{e1}
\end{equation}

Since $R_\|(\lambda)$ and $R_\bot(\lambda)$ are complicated functions
of $n$ and $k$, so are $\Plam$ and $\Dlam$.  The equations for $\Plam$
and $\Dlam$ in terms of $n$ and $k$ can not be algebraically inverted
to give $n$ and $k$ in terms of $\Plam$ and $\Dlam$ except under the
simplest of circumstances.  Thus, the determination of $\nlam$ and
$\klam$ (and the sample thickness) from $\Plam$ and $\Dlam$ measured
by the ellipsometer is a problem which requires a numerical solution.

We developed a non-linear least-squares fitting technique using
MATHMATICA to determine $\nlam$, $\klam$ and the film thickness for
this HAC sample.  In our method, $\nlam$ and $\klam$ were assumed to
be parameterized functions of wavelength, and the values of the
parameters were determined so as to minimize the difference between
the $\Plam$ and $\Dlam$ measured at a given angle of incidence and
those computed using the parameterized $\nlam$ and $\klam$ functions.
The thickness was also a parameter adjusted during the fitting
process.  The general form for the parameterized functions describing
$\nlam$ and $\klam$ were determined by trial and error fitting to $n$
and $k$ data presented by Smith (1984)\markcite{s84} for a number of
HAC samples ranging from highly hydrogenated and transparent to
graphitic and more opaque.  It was found that the following functional
forms for $\nlam$ and $\klam$ were the simplest that could fit, given
the choice of appropriate parameters, all of the data given by Smith
(1984)\markcite{s84}:

\begin{eqnarray}
\nlam & = & n_o + n_1\lambda^{-1} + n_2\lambda^{-2} \;\;, \label{e2}\\
\klam & = & k_o \; exp (k_1\lambda + k_2\lambda^{-1})\;\;. \label{e3}
\end{eqnarray}

\begin{figure*}
\plotone{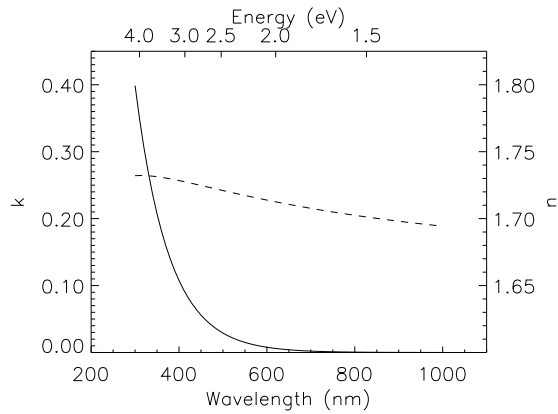}
\caption{The components of the complex index of refraction ($n - ik$)
for the interstellar HAC analog.  The dashed line is $n$, the solid
line is $k$. \label{f3}}
\end{figure*}

Values for the parameters describing $\nlam$ and $\klam$ that
minimized the least squares difference between measured and computed
$\Plam$ and $\Dlam$ were determined for the HAC sample and for the
fused silica substrate itself at four angles of incidence evenly
spaced between $60^\circ\!-\!70^\circ$.  The values of $n$ and $k$
derived for the fused silica blank agreed well with literature values,
thus verifying the technique, and it was found for the HAC sample that
one set of parameters could reproduce all the experimental data to a
level of $\chi^2 \sim 10^{-3}$.  The solutions were also found to be
unique.  The functions that emerged as solutions were generally
insensitive to the initial guesses for the parameters, and were
unchanged by the addition of other parameters (i.e., $n_3$, $n_4$,
\ldots, $k_3$, $k_4$, \ldots).  The values for the parameters of
Equations~\ref{e2} and~\ref{e3} describing $\nlam$ and $\klam$ for
this HAC sample are shown in Table~\ref{t2}; the thickness of the film
was found to be $300 \pm 30\;nm$.  Plots of $\nlam$ and $\klam$ are
shown in Figure~\ref{f3}.

\begin{deluxetable}{cccc}
\tablecaption{Interstellar HAC Analog Index of Refraction
Parameters \label{t2}}
\tablehead{\colhead{Parameter} & \colhead{Value} & \colhead{Parameter}
& \colhead{Value}}
\startdata
$n_o$ & 1.652 & $k_o$ & 20.3 \nl
$n_1$ & $50.2\;nm$ & $k_1$ & $-0.0131\;nm^{-1}$ \nl
$n_2$ & $-7854\;nm^2$ & $k_2$ & 0 \nl
\enddata
\tablecomments{Values of the parameters giving $\nlam$ and $\klam$ 
via Equations~\ref{e2} and~\ref{e3} for the interstellar HAC analog.  
Valid wavelength range is $300 \leq \lambda \leq 1000\;nm$; sample 
thickness is $300 \pm 30\;nm$.}
\end{deluxetable}

It should be noted that our determinations of $\nlam$ and $\klam$ are
certain to approximately $10\!-\!20\%$ at any given wavelength, and
that they compare favorably to other published data for similar HAC
materials (e.g., Smith 1984\markcite{s84}; Alterovitz
\etal~1991\markcite{a91}) and for organic grain mantle materials
(Chlewicki \& Greenberg 1990\markcite{cg90}; Jenniskens
1993\markcite{j93}).

\subsection{HAC Sample IR Mass-extinction Coefficient}

\begin{figure*}
\plotone{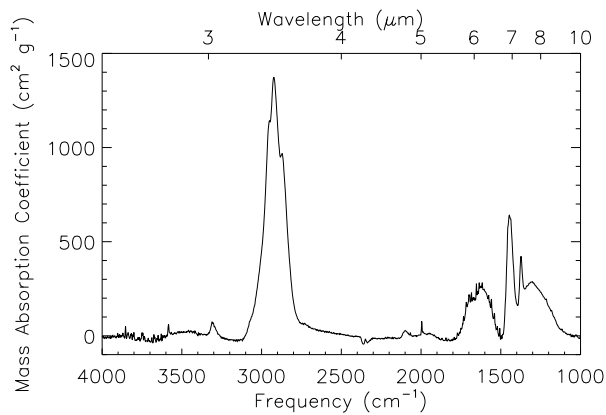}
\caption{IR mass-absorption coefficient for the interstellar HAC
analog derived as described in the text.\label{f4}}
\end{figure*}

The IR absorption spectrum of this interstellar HAC analog was 
recorded over the frequency range $4000\!-\!1000\;\icm$ with a Nicolet 
510P FT-IR spectrometer.  The sample was deposited onto a NaCl 
substrate for this analysis.  The resulting spectrum, percent 
transmission as a function of frequency, was quantified by reducing it 
to spectral mass-absorption coefficient as follows.  First, the raw 
absorption spectrum was divided by an absorption spectrum of the NaCl 
substrate itself and baseline corrected.  This was done in order to 
highlight the principle absorption features, and to suppress continuum 
absorption by the HAC which is primarily due to scattering and 
interference from the HAC/NaCl interface and the exposed surfaces.  
The baseline-corrected spectrum was then converted to optical 
depth and reduced to mass-absorption according to the relation

\begin{equation}
\kappa (\nu) = \frac{-ln(I/I_o)}{\rho t}\;\;, \label{e4}
\end{equation} 

where $\rho$ and $t$ are the sample density and thickness determined
as previously described.  The final spectrum is shown in
Figure~\ref{f4}.  The small, double feature at $\sim \! 2350\;\icm$
and the high-frequency structure in the $2000\!-\!1500\;\icm$ range
and above $\sim\!\!3500\;\icm$ are not associated with the HAC sample;
they arise from water vapor and CO$_2$ not completely purged from the
spectrometer when the spectrum was recorded.

Inspection of this plot reveals that the single, strongest absorption
feature in this type of HAC is the $\chcm$ C-H stretching band, with a
maximum mass-absorption coefficient of $1.4\EE{3}\;cm^2/g$.  Other
absorption features are also apparent at the low-frequency end of the
spectrum, but at levels significantly weaker than the main C-H
stretching peak; these bands are due to C-H wagging and C-C stretching
modes.

The integrated absorption strength for 
this sample in the $\chcm$ feature is $\kappa = 46\;cm^{2}/g$, which 
corresponds to an intrinsic, integrated band strength of $\sigma = 
1.7\EE{-21}\;cm^{2}$ per C-H bond.\footnote{See \S3.1 for the details of 
this calcualtion.}  From Figure~1 in Jacob \& Unger 
(1996)\markcite{ju96} we thus find that H/C for this sample is about 
0.5 as presented in \S2.2.  It should be pointed out that only bound 
hydrogen is included in this ratio.

\section{Discussion}

Carbon is the most abundant element capable of maintaining itself in
solid form under the conditions that prevail in the diffuse ISM.
Various forms of solid carbon are invoked as sources or carriers of a
variety of phenomena; for example, graphite and amorphous carbon
grains for the $220\;nm$ bump (Mathis 1994\markcite{m94}; Mennella
\etal~1998\markcite{m98}), HAC grains for ERE and for near-IR
continuum emission (Duley 1985\markcite{d85}), and gas-phase HAC
fragments for diffuse interstellar bands (Duley 1995\markcite{d95}).
However, the only direct evidence that some form of solid carbonaceous
material is present in the diffuse ISM is the detection of the $\chcm$
absorption feature due to C-H bonds along dusty lines of sight (e.g.,
Soifer, Russell, \& Merrill 1976\markcite{srm76}; Adamson, Whittet, \&
Duley 1990\markcite{awd90}; Sandford \etal~1991\markcite{s91};
Pendleton \etal~1994\markcite{p94}; Whittet \etal~1997\markcite{w97}).
This detection and identification are firm.  The extent to which HAC
can account for other astrophysical phenomena depends on its optical
and physical properties, as derived from laboratory investigations of
these materials.

\begin{figure*}
\plotone{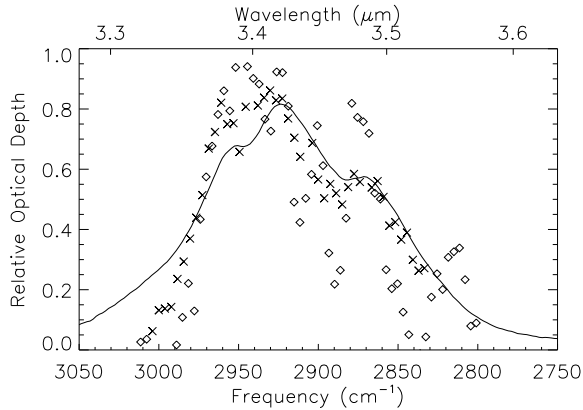}
\caption{The $2750\!-\!3050\;\icm$ relative optical depth spectrum for
the interstellar HAC analog, along with that observed toward the
Galactic center source IRS 6 by Pendleton \etal~(1994)
($\times\!\!\times\!\!\times$) and toward Cyg.~OB2~\#12 by
Whittet \etal~(1997) using ISO ($\Diamond \Diamond \Diamond$). \label{f5}}
\end{figure*}

The most superficial comparison that can be made between
laboratory-produced HAC and the true interstellar HAC, but nonetheless
an important one, is of the profile of the C-H stretch feature.
Indeed, comparisons of this sort are plentiful in the literature
(e.g., Sandford \etal~1991\markcite{s91}; Duley 1994\markcite{d94};
Greenberg \etal~1995\markcite{g95}; Schnaiter 
\etal~1998\markcite{s98}).  We provide our own such comparison in
Figure~\ref{f5}, which shows the $2750\!-\!3050\;\icm$ relative
optical depth spectrum for the interstellar HAC analog presented here,
along with that observed toward the Galactic center source IRS 6 and
toward Cyg.~OB2~\#12.  The IRS 6 data were provided by Y.~Pendleton
and are published in Pendleton \etal~(1994)\markcite{p94} the
Cyg.~OB2~\#12 data recorded by ISO were provided by D.C.B.~Whittet and
are published in Whittet \etal~(1997)\markcite{w97}.  It is clear from
this figure that the two interstellar profiles are similar and are
both well represented by the profile of the interstellar HAC analog
(see also Chiar \etal~(1998)\markcite{c98} for additional, very recent
comparisons of this sort).  Principal uncertainties in the data are
associated with determining the appropriate baseline to use to extract
the profile of the feature from the astronomical spectra.

It is, however, not sufficient to identify a laboratory-produced
material as an analog to some interstellar material based on the
observation of a single absorption feature.  What other absorption
features are produced by dust in the diffuse ISM?  This question is
most unambiguously answered by the recent observation of Cyg.~OB2~\#12
by Whittet \etal~(1997)\markcite{w97} using ISO.  Absorption along
this line of sight is believed to arise nearly entirely from dust in
the diffuse ISM.\footnote{At least some of the absorption along the
lines of sight to IR-bright Galactic center sources such as IRS~6 is
known to arise from molecular cloud edges (McFadzean
\etal~1989\markcite{m89}).} Perhaps the most striking fact about the
ISO Cyg.~OB2~\#12 observation is the lack of any strong absorption
features in addition to the Si-O band at $\sim\!\!1000\;\icm$ and the
C-H band at $\chcm$, except possibly a very weak feature at
$\sim\!\!1600\;\icm$, either due to C-C stretching or C-H wagging
vibrations.  This, however, is useful evidence in itself which serves
to exclude a large number of interstellar carbonaceous grain analog
materials that have been proposed to date.  Note, in this respect,
that the IR absorption spectrum of the interstellar HAC analog
presented here (Figure~\ref{f4}) is dominated by C-H stretching
absorption at $\chcm$, with much less prominent absorption in the
$1000\!-\!1800\;\icm$ region which would be still less conspicuous at
the signal-to-noise of the ISO Cyg.~OB2~\#12 observation.

In addition to comparing the profiles of IR absorption features
observed along lines of sight through the diffuse ISM with those
produced by laboratory analogs, it is possible and indeed necessary to
address the question of absorption-band strength.  How much carbon
needs to be locked in the form of interstellar HAC in order to explain
the {\em strength} of the interstellar $\chcm$ feature?  This question
is crucial in light of the recent determinations of the amount of
solid carbon likely to be available to form interstellar HAC material
(Snow \& Witt 1995\markcite{sw95}).

\subsection{The Amount of Carbon in Diffuse ISM HAC Grains}

In principle it is straight forward to determine the amount of carbon
locked in diffuse ISM HAC grains from a quantitative analysis of the
$\chcm$ interstellar absorption feature due to C-H bonds.  One
measures the total optical depth in the band (either by integrating
over the band profile or by simply noting the maximum optical depth of
the feature) and determines the number of C-H bonds necessary to
produce the absorption using laboratory measurements of the intrinsic
absorption strength or cross section per C-H bond.  Indeed, several
groups have done this, most notably Sandford \etal~(1991)\markcite{s91}
(hereafter S91) and Duley \etal~(1998)\markcite{d98} (hereafter D98).
But, in practice, there are significant difficulties with the laboratory
measurement and interpretation of the C-H stretching absorption-band
strength which have been overlooked.

S91 analyzed the $\chcm$ C-H absorption feature observed toward
several bright IR sources believed to be probes of the diffuse ISM.
They concluded that the amount of carbon necessary to produce the
observed optical depth is $\sim\!\!10\%$, to within a factor of two,
of the cosmic abundance of carbon as presented in Allen 
(1973)\markcite{a73} ($370\;ppM$ H), or roughly $20\!-\!80\;ppM$ of 
hydrogen.  They relied, however, on C-H absorption-band strengths (and 
feature widths) measured in small hydrocarbon molecules 
(C$_5$H$_{12}$, C$_6$H$_{14}$, and cyclo-C$_6$H$_{14}$), when there is 
no {\em a priori} reason to expect the C-H stretching band strength in 
small molecules to be the same as that in solid matter.  D98 measured 
the C-H absorption band strength for a single HAC produced by laser 
ablation of carbon in a low pressure, hydrogen-rich atmosphere and 
analyzed the IR observations of lines of sight toward the Galactic 
center by Pendleton \etal~(1994)\markcite{p94} to conclude that 
$72\!-\!97\;ppM$ H of carbon is necessary to explain the optical depth 
in the interstellar $\chcm$ feature.  But here, the assumption is made 
that the C-H stretching band strength is independent of the H/C ratio 
in HAC, when there is in fact evidence to the contrary (Jacob \& Unger 
1996\markcite{ju96}, hereafter JU96).  It is clear from both of these 
studies that the laboratory determinations of the intrinsic absorption 
strengths of various C-H modes contribute a large uncertainty to the 
determination of the amount of carbon locked in interstellar HAC.
 
The literature concerning the quantitative analysis of IR absorption 
features is especially confusing to the uninitiated; there are about 
as many units for characterizing the intrinsic strength of an IR mode 
as there are papers doing so.  The following paragraphs serve to 
connect two frameworks relevant to this current work.

The relation most commonly used in the astronomical literature, and 
the one used by S91, among others is

\begin{equation}
N = \frac{1}{A} \; \tau_{max} \; \Delta \nu \;\;, \label{e5}
\end{equation}

\noindent where N is the column density of absorbers, $\tau_{max}$ is 
the maximum optical depth in an IR absorption feature and $\Delta \nu$ 
is the full-width at half maximum of the feature in wavenumbers.  $A$ 
is a constant with units $cm\;per\;group$ or $cm\;per\;bond$.  
Equation~\ref{e5} is an approximate relation, derived from the more 
formal, rigorous definition for the number density of C-H bonds,

\begin{equation}
n = \frac{1}{\sigma} \int_{feature} \frac{\alpha(\nu)}{\nu} \; d\nu \;\;, \label{e6}
\end{equation}

\noindent where $\alpha(\nu)$ is the absorption coefficient as a 
function of frequency and the normalization constant $\sigma$ is the 
integrated cross section per absorber, in units $cm^2 \; per \; 
absorber$.  Some physical-chemistry studies of IR band strengths 
report band cross sections, not the $A$-value of Equation~\ref{e5}, 
most notably in this context the work of JU96.  Note that there is an 
approximate relation between $A$ and $\sigma$

\begin{equation}
A \simeq \sigma \bar{\nu} \;\;, \label{e7}
\end{equation}

\noindent where $\bar{\nu}$ is the average or central frequency of the
absorption band.

In connection with determining the amount of carbon locked in 
interstellar HAC, the work of JU96 is of utmost relevance.  These 
authors determined the integrated absorption cross section per C-H 
bond experimentally in a large number of HAC samples covering a range 
in H/C ratio from about 0.3 to about 1.\footnote{They define a 
constant $A$ which is not the Sandford \etal\ $A$, but that is the 
reciprocal of the integrated band cross section.} They found that the 
cross section varied linearly with H/C such that $1 \EE{-21} \leq 
\sigma \leq 5 \EE{-21}\;cm^2$ for $0.3 \leq \; $H/C$ \; \leq 
1.$\footnote{There is some scatter in the data presented by JU96, and 
possibly an indication of a saturation of the $\chcm$ band strength at 
high levels of hydrogenation, but these attributes do not 
significantly affect the discussion and conclusions that follow.}
This result complicates the analysis of the interstellar $\chcm$ 
feature because it reveals that the H/C ratio in interstellar HAC 
grains must be estimated in order to determine how much carbon these 
grains contain, since the intrinsic band strength is a function of the 
H/C ratio.

Before proceeding to discuss the results of JU96 more thoroughly, it 
is instructive to compare their results to those of S91 and D98.  Both 
of these groups determined $A$-values for four principal $sp^3$ C-H 
stretching modes: --CH$_3$ symmetric and asymmetric, and $>$CH$_2$ 
symmetric and asymmetric.  Their results are summarized in 
Table~\ref{t3}; $A$-values were converted to cross sections using 
Equation~\ref{e7}.

\begin{deluxetable}{lccccc}
\tablecaption{
Summary of Measured C-H Stretch Absorption Band Strengths \label{t3}
}
\tablehead{
\colhead{}          & \colhead{$\nu$}          & \colhead{$A_{S91}$} & 
\colhead{$A_{D98}$} & \colhead{$\sigma_{S91}$} & \colhead{$\sigma_{D98}$} \\
\colhead{Mode}      & \colhead{($cm^{-1}$)}    & \multicolumn{2}{c}{($10^{-18}\;cm/group$)} & \multicolumn{2}{c}{($10^{-21}\;cm^2/bond$)}
}
\startdata
--CH$_3$ asy. & 2955 & 12 & 1.7 & 1.4 & 0.19 \nl
$>$CH$_2$ asy. & 2925 & 7.4 & 2.6 & 1.3 & 0.44 \nl
-CH$_3$ sym. & 2870 & 2.1 & 2.3 & 0.24 & 0.27 \nl
$>$CH$_2$ sym. & 2955 & 2.1 & \nodata & 0.37 & \nodata \nl
  &  &  & \multicolumn{1}{r}{Total:} & 3.3 & 0.90 
\enddata
\tablecomments{Summary of measured ``$A$-values'' for the C-H
stretch band from S91 and D98, along with integrated band cross
sections derived from these data using Equation~\ref{e7}.  Cross
sections determined by JU96 for HAC materials ranged linearly between
$1 \EE{-21} \leq \sigma \leq 5 \EE{-21}\;cm^2$ for $0.3 \leq \; $H/C$
\; \leq 1$ }
\end{deluxetable}

The results in Table~\ref{t3} are consistent with those of JU96.  The
relatively low value of $\sigma_{D98}$ is consistent with their
determination of H/C in the range $0.3\!-\!0.4$ for the sample.  The
relatively high value of $\sigma_{S91}$ is consistent with the fact
that the hydrocarbon molecules they analyzed were saturated, with H/C
in excess of unity.  The interstellar HAC analog presented in our work
has $H/C \simeq 0.5$ and we determine the integrated absorption cross
section, according to Equation~\ref{e6}, to be $\sigma =
1.7\EE{-21}\;cm^2$, which is also consistent with the results of JU96.

These results clearly demonstrate the problem which complicates the
analysis of the interstellar $\chcm$ feature: the intrinsic cross
section per C-H bond depends strongly on the H/C ratio of the
interstellar HAC grains, which is not well determined.  The only data
that give an indication of the H/C ratio in the interstellar material
is the profile of the C-H stretch feature itself.  Materials with very
low hydrogen content (and very low $sp^3/sp^2$ ratios) have profiles
dominated by $sp^2$-modes which occur at frequencies near and above
$3000\;\icm$; materials with high hydrogen content (and higher
$sp^3/sp^2$ ratios) have profiles dominated by $sp^3$-modes, as seems
to be the astrophysical situation.  But the profile is not very
sensitive for HACs with H/C greater than about 0.3, perhaps because
there is a preference for hydrogen to bond to $sp^3$-hybridized
carbon.  So, the profile of the C-H stretch feature alone is not
sufficient to remove the complication.

The mass-absorption coefficient $\kappa$ for the C-H stretch feature 
is the quantity that directly relates the strength of the observed 
interstellar band to the amount of carbon locked in interstellar HAC 
grains.  $\kappa$ is defined to be the cross section per unit mass of 
material

\begin{equation}
\kappa = \frac{\sigma}{m_{CH}} \;\; , \label{e8}
\end{equation}

\noindent where $m_{CH}$ is the mass of material per C-H bond since
$\sigma$ is the cross section per C-H bond.  But, the definition

\begin{equation}
\kappa^{\prime}(\nu) = \frac{\alpha(\nu)}{\rho} \;\; , \label{e9}
\end{equation}

\noindent where $\alpha(\nu)$ is the absorption coefficient as a 
function of frequency and $\rho$ is the mass-density, is also commonly 
used.  These equations define $\kappa$'s which are different but 
related -- as reviewed below.  

Equation~\ref{e6}, which rigorously defines $\kappa$, can be 
approximated by

\begin{equation}
n \simeq \frac{1}{\sigma} \; \alpha_{max} \; \frac{\Delta \nu}{\bar{\nu}}
\;\; , \label{e10}
\end{equation}

\noindent where $\alpha_{max}$ is the maximum absorption coefficient
in the feature, and $\Delta \nu$ and $\bar{\nu}$ are the feature width
and central frequency, respectively.  For the interstellar C-H
feature, $\Delta \nu / \bar{\nu} \simeq  \frac{1}{30}$.  Given that $n
= \rho / m_{CH}$ is the number-density of C-H bonds, combining
Equations~\ref{e9} and~\ref{e10} gives

\begin{equation}
\kappa \simeq \frac{1}{30} \; \kappa^{\prime}(2950) \;\; \label{e11}
\end{equation}

\noindent as the relation between $\kappa$ determined from $\sigma$
according to Equation~\ref{e8} and $\kappa^{\prime}$ at the
interstellar $\chcm$ feature maximum defined by Equation~\ref{e9}.

Thus Equation~\ref{e11} can be used to relate $\kappa$, the integrated 
mass-absorption coefficient, to $\kappa^{\prime}(2950)$ which is most 
commonly used in the astronomy literature.

The mass-absorption coefficient for the C-H stretch feature in HAC
materials depends on the H/C ratio.  In Equation~\ref{e8}, both $\sigma$
and $m_{CH}$ are functions of the H/C ratio; specifically, it can be
shown that 

\begin{equation}
m_{CH} = m_H + \frac{m_C}{{\rm H/C}} \;\; , \label{e12}
\end{equation}

\noindent where $m_H$ and $m_C$ are the mass of the hydrogen and
carbon atoms, respectively.  Using this relation, the values for
$\sigma$ determined by JU96, and Equation~\ref{e11}, the 
mass-absorption coefficient as a function of H/C ratio can be 
computed as presented in Table~\ref{t4}.  It is clear from these
data that the mass-absorption coefficient for the C-H band in HAC
materials varies by over a factor of 20 for $0.2 \leq \; $H/C$ \; 
\leq 1$

\begin{deluxetable}{ccccc}
\tablecaption{C-H Stretch Mass-absorption Coefficient for HACs \label{t4}}
\tablehead{
\colhead{H/C} & \colhead{$\sigma$} & \colhead{$m_{CH}$} & 
\colhead{$\kappa$} & \colhead{$\kappa^{\prime}$} \\
\colhead{}         & \colhead{($10^{-21}\;cm^2$)} & 
\colhead{($10^{-23}\;g$)} & \colhead{($cm^2/g$)} & \colhead{($10^3\;cm^2/g$)}
}
\startdata
1.00 & 5.0 & 2.2 & 230 & 6.9 \nl
0.80 & 3.2 & 2.7 & 120 & 3.6 \nl
0.67 & 2.0 & 3.1 & 64  & 1.9 \nl
0.33 & 1.3 & 6.1 & 21  & 0.63 \nl
0.20 & 1.0 & 10. & 10. & 0.30 
\enddata
\tablecomments{The values for $\sigma$ in Column~2 are from Jacob \&
Unger (1996)\markcite{ju96}.}
\end{deluxetable}

What is the mass-absorption coefficient for interstellar HAC grains as
determined observationally? The mass-absorption coefficient
$\kappa^{\prime}$ defined by Equation~\ref{e9} can be thought of as
the optical depth per the column mass-density of absorbers.
Accordingly, it can be stated in terms of relevant astrophysical
quantities as follows:

\begin{equation}
\kappa^{\prime}(2950) = \frac{\tau(2950)}{N({\rm H})} 
\left( \frac{N({\rm C})}{N({\rm H})} \right)^{-1}_{\delta HAC} \! \!
\left( \frac{{\rm C}/{\rm H}}{m_{CH}} \right) \;\; , \label{e13}
\end{equation}

\noindent where $N({\rm X})$ is column density of element $X$, $\left(
\frac{N({\rm C})}{N({\rm H})} \right)^{-1}_{\delta HAC}$ is the amount
of carbon relative to hydrogen depleted into the HAC grains (the
quantity we wish to determine), and C/H is that of the HAC material
itself.

In what follows, we will use Equation~\ref{e13} to analyze exclusively
the line of sight toward Cyg.~OB2~\#12, the prototypical diffuse ISM
line of sight.  Similar results are obtained for lines of sight toward
the Galactic center.  The maximum optical depth in the $\chcm$ feature
observed toward Cyg.~OB2~\#12 is $\tau(2950) \simeq 0.04$ with $A_V =
10.2\;mag$.  Using $\tau_V = 4.92\EE{-22}\;N({\rm H})$ given by Mathis
(1996)\markcite{m96}, we find $\tau(2950)/N({\rm H}) =
2.1\EE{-24}\;cm^{-2}$ (which is consistent with other lines of sight
as well).  The ratio $\left( \frac{{\rm C}/{\rm H}}{m_{CH}} \right)$,
derived from the values in Table~\ref{t4}, ranges from
$4.5\EE{22}\;g^{-1}$ to $5.0\EE{22}\;g^{-1}$, so a constant value of
$4.7\EE{22}\;g^{-1}$ is representative of HACs to within the
uncertainty of the observations.  Thus, the mass-absorption
coefficient for the interstellar C-H stretch feature is

\begin{equation}
\kappa^{\prime}(2950) = \left( 0.10\;cm^2\;g^{-1} \right)
\left( \frac{N({\rm C})}{N({\rm H})} \right)^{-1}_{\delta HAC} \label{e14}
\end{equation}

\noindent which depends only on the amount of carbon relative to
hydrogen depleted into HAC grains.  Notice that the system of
variables: $\left( \frac{N({\rm C})}{N({\rm H})} \right)_{\delta
HAC}$, H/C for HAC, and $\kappa^{\prime}(2950)$ is underconstrained.
There are three variables, but only two constraints: the relation
between H/C for HAC and $\kappa^{\prime}(2950)$ summarized in
Table~\ref{t4}, and the relation between $\left( \frac{N({\rm
C})}{N({\rm H})} \right)_{\delta HAC}$ and $\kappa^{\prime}(2950)$
given in Equation~\ref{e14}.  Table~\ref{t5} summarizes this.  The
first column lists values for the amount of carbon relative to
hydrogen locked in interstellar HAC grains $\left( \frac{N({\rm
C})}{N({\rm H})} \right)_{\delta HAC}$; the second column shows the
corresponding, required values of $\kappa^{\prime}(2950)$.  The third
column gives values for the HAC H/C ratio associated with the
$\kappa$'s in the second column, interpolated from Table~\ref{t4}.
Note the approximate inverse relationship between the amount of carbon
locked in HAC grains and the H/C ratio the HAC material would need to
have in order to explain the observed optical depth in the
interstellar $\chcm$ feature.

\begin{deluxetable}{ccc}
\tablewidth{4.5in}
\tablecaption{The Amount of Carbon in Interstellar HAC \label{t5}}
\tablehead{
\colhead{$\left( \frac{N({\rm C})}{N({\rm H})}\right)_{\delta HAC}$} & 
\colhead{$\kappa^{\prime}(2950)$} & \colhead{H/C} \\
\colhead{($ppM$)} & \colhead{($10^3\;cm^2/g$)} & \colhead{}
}
\startdata
100 & 1.0 & 0.4 \nl
80  & 1.3 & 0.5 \nl
60  & 1.7 & 0.6 \nl
40  & 2.5 & 0.7 \nl
20  & 5.0 & 0.9 \nl
\enddata 
\end{deluxetable}

Thus we are lead to the conclusion that the amount of carbon locked in 
interstellar HAC grains is in the range of $80\;ppM$ hydrogen (or 
more, depending on the gas-phase abundance and depletion of carbon) to 
possibly less than $20\;ppM$ hydrogen, depending on the degree of 
hydrogenation of the material.  Note that this result is similar to 
that presented by S91 with one significant difference: the large range 
in their estimate of $\left( \frac{N({\rm C})}{N({\rm H})} 
\right)_{\delta HAC}$ is, by their own acknowledgment, due to 
uncertainties in the observations and in band-strengths they rely on.  
This is not the case here; the large range in $\left( \frac{N({\rm 
C})}{N({\rm H})} \right)_{\delta HAC}$ is due to a fundamental lack of 
information to close the problem.

We can, however, draw important conclusions from this analysis, and
can use these results to place limits on the interstellar HAC grain
problem.  On the surface, it is clear that interstellar HAC grains must be
hydrogenated at least to the level H/C$\;\simeq 0.5$, and at this
limit most of the available solid carbon would need to be in
form of HAC grains.  The other limit is not so precisely
delineated.  It is clear that if the HAC grains are extremely highly
hydrogenated, to the maximum H/C$\;\simeq 1\!-\!1.5$, then the amount
of carbon needed to produce the observed optical depth in the $\chcm$
feature would be small, perhaps as low as $20\;ppM$ hydrogen.
But, as we discuss in \S3.3, this seems unlikely because these HACs
photoluminesce efficiently in the blue-green and such emission has
never been observed in interstellar environments.

\subsection{Viable Interstellar HAC Analogs}

The data in Table~\ref{t5} can be used to evaluate the viability of
other potential interstellar HAC analogs.  From a purely qualitative
standpoint, any material proposed as an analog to the hydrogenated
carbonaceous component of the diffuse ISM must produce absorption in
the IR that is dominated by the C-H stretch feature at $\chcm$, given
the recent ISO observation of Cyg.~OB2~\#12 by Whittet
\etal~(1998)\markcite{w98}.  The lack of other strong features along
this line of sight (excepting the Si-O band at $1000\;\icm$) is
meaningful.  In light of this, the suggestion that solids such as the
organic refractory EUREKA materials of Greenberg
\etal~(1995)\markcite{g95}, for example, are analogs to the
carbonaceous component of the diffuse ISM must not be taken too
literally because they show significant absorption due to bonds other
than C-H and C-C.  These materials do, however, offer intriguing
insight into a possible interstellar HAC production method.

From a quantitative standpoint, the results of the previous section
provide a key criterion to which all interstellar HAC analogs must be
held.  Given that only $80\;ppM$ of carbon relative to hydrogen is
likely to be depleted from the gas phase to form solid carbonaceous
material (Snow \& Witt 1995\markcite{sw95}), any laboratory analog to
the interstellar HAC material must have a mass-absorption coefficient
at the peak of the C-H stretch feature in excess of $\sim
10^3\;cm^2/g$, and in order to do so must be hydrogenated to a level
of H/C$\;> 0.5$.

The interstellar HAC analog presented in this work, with H/C $\simeq 
0.5$ and $\kappa^{\prime}(2950) = 1.4\EE{3}\;cm^2/g$, appears to be a 
limiting material in the sense that it produces absorption in the C-H 
stretch feature with a strength that would require all of the 
available solid carbon to be in form of this type of material.  Other 
interstellar HAC analogs presented to date, however, appear in this 
analysis to require more the solid carbon than is available in order 
to account for the interstellar $\chcm$ feature (Borghesi 
\etal~1985\markcite{b85}; Ogmen \& Duley 1988\markcite{od88}; 
Colangeli \etal~1995\markcite{c95}; Jaeger \etal~1998\markcite{j98}).
Of course, uncertainty in the true amount of 
solid-phase carbon in the ISM constrains each of these interstellar 
HAC analogs to different degrees; each must be considered on a case by 
case basis.  However, that materials with lower mass-absorption 
coefficients in the $\chcm$ band require more solid-phase carbon to 
account for the observed strength of this feature -- perhaps more than 
is available -- remains a valid conclusion.

While determination of IR mass-absorption spectra for interstellar 
grain analogs is a necessary quantification, it has only been done by 
a limited number of groups.  Bussoletti \etal~(1987)\markcite{b87} and 
Colangeli \etal~(1995)\markcite{c95} produced and analyzed a variety 
of HAC materials (denoted BE, ACAR, ACH2, etc\ldots), derived by 
arc-evaporation of graphite, and determined mass-absorption spectra 
for them through the near-IR along with other properties, including 
indices of refraction, which, incidently, have been used subsequently 
by Mathis (1996)\markcite{m96} to compute his most recent dust model.  
For these materials, however, $\kappa^{\prime}(2950)$ is only on the 
order of $100\!-\!150\;cm^2/g$ -- clearly too low to account for the 
interstellar $\chcm$ feature.  More recently, Schnaiter 
\etal~(1998)\markcite{s98} produced a variety of matrix-isolated 
nano-sized carbon grains and, among other things, determined the 
mass-absorption coefficient in the C-H stretch feature.  They found 
$\kappa^{\prime}(2950)$ to range from near zero to around 
$600\;cm^2/g$ for the most highly hydrogenated materials they 
produced.  The mostly highly hydrogenated of 
these materials may possibly be able to account for the $\chcm$ 
feature, given the uncertainty in the true amount of solid-phase 
carbon that is available.  Finally, D98 produced an interstellar HAC 
analog by laser ablation of graphite in a low-pressure, hydrogen-rich 
atmosphere, and analyzed the strength of the C-H stretch feature.  
Although they do not specifically compute $\kappa^{\prime}(2950)$, it 
is apparently approximately $450\;cm^2/g$, based on the H/C ratio they 
quote and the ``$A$-value'' (of Equation~\ref{e5}) they determine; 
again considerably lower than necessary to account for all of the 
absorption in the interstellar $\chcm$ feature.  So, although a number 
of carbonaceous materials have been presented in the literature over 
the past few years as sources of the interstellar $\chcm$ feature, and 
thus as interstellar HAC analogs, the HAC presented in this work is 
the first to be shown to be {\em quantitatively} consistent with IR 
observations in the $4000\!-\!1000\;\icm$ spectral region of lines of 
sight which unambiguously sample only the diffuse ISM.

\subsection{The Optical Role of HAC Grains in the Diffuse ISM}

Given the presence in the diffuse ISM of some form of hydrogenated
carbonaceous solid, as betrayed by the interstellar $\chcm$ feature,
it is necessary to consider carefully the roles it may and may not
play in other UV/visible/IR phenomena.  HAC is included as a source of
opacity in the current dust models of Mathis (1996)\markcite{m96} and
Li \& Greenberg (1997)\markcite{lg97}, and has been proposed as the
source or carrier of the broad, red dust luminescence band known as
ERE (e.g., Duley 1985\markcite{d85}; Witt 1994\markcite{w94}), near-IR
($1\!-\!3\;\mu m$) non-thermal continuum radiation (Jones, Duley, \&
Williams 1990\markcite{jdw90}), and the $220\;nm$ bump (Colangeli
\etal~1993\markcite{c93}; Schnaiter \etal~1998\markcite{s98}), in
addition to being the assumed source of the interstellar $\chcm$ feature. 
The extent to which HAC grains can indeed fill all of these roles must
continue to be reviewed in light of new astronomical and laboratory
data.

Although HAC is assumed to be a major source of opacity in the near-UV 
and visible (Li \& Greenberg 1997\markcite{lg97}), its optical 
constants in this spectral region are by far the poorest determined of 
most astrophysically important materials.  In addition, the recent 
recent dust models of Li \& Greenberg 1997\markcite{lg97} and Mathis 
(1996)\markcite{m96} require that the imaginary part of the index of 
refraction for the material causing visible/near-UV absorption needs 
to be as high as $0.4$.  If so, then the material causing this 
absorption is not the HAC responsible for the interstellar $\chcm$ 
feature, for the simple reason that HACs which produce significant 
optical depth per carbon atom in the C-H stretch band are not nearly 
this absorbing in the visible.  As demonstrated in Figure~\ref{f3}, 
our HAC sample, which matches the interstellar C-H feature both in 
profile and in mass-absorption coefficient (at a minimum level), is 
essentially transparent in the visible, with $k$ rising to about $0.4$ 
only in the near-UV. Further, we consider this a limiting material 
because increasing its visual opacity would require lowering its 
bandgap below its present value ($1.9\;eV$), decreasing its 
mass-absorpton coefficient in the C-H stretch band to where more than 
$80\;ppM$ solid phase carbon would be required to match the C-H band 
strength, and shifting its C-H band profile toward frequencies higher 
than observed.  HAC, then, is not the sole source of absorption in the 
visible/near-UV.

A recent paper by Adamson \etal~(1998)\markcite{a98}, describing 
spectropolarimetric observations of the interstellar $\chcm$ feature 
observed toward the Galactic center source Sgr.~A~IRS7, provides an 
additional clue to the nature of the carrier of this absorption band.  
They find this feature to be unpolarized with an upper limit well 
below that expected on the basis of a model in which the carrier bonds 
are associated with the aligned silicate grains (i.e., 
silicate-core/HAC-mantle grains).  This suggests the possibility that 
the IR-active HAC grains are extremely small -- below the Rayleigh 
limit.  If this is the case, then it seems again that the HAC material 
is very highly hydrogenated with a correspondingly high C-H stretch 
mass-absorption coefficient.  Interestingly enough, small HAC grains 
have recently been shown in the lab by Schnaiter 
\etal~(1998)\markcite{s98} to produce broad absorption around 
$220\;nm$, thus making them candidates for the carrier of the 
interstellar $220\;nm$ extinction feature.  They concluded, in fact, 
that small HAC grains locking up about $100\;ppM$ carbon relative to 
hydrogen could explain the observed strength of the $220\;nm$ bump. 
In addition, these laboratory-produced HAC 
grains have a mass-absorption coefficient in the $\chcm$feature of 
$\kappa^{\prime}(2950) \simeq 600\;cm^2/g$, which is near but lower 
than the required $10^{3}\;cm^2/g$.  It thus appears that small HAC 
grains are a promising source of both the $220\;nm$ bump and the 
$\chcm$ C-H feature.  But, they would need to lock up essentially all 
the solid-phase carbon if the recent works of Schnaiter 
\etal~1998\markcite{s98} and Schnaiter \etal~1999~\markcite{s99} are 
final.  It is important to note that the HAC sample presented in this 
current work does not produce structure in the $190\!-\!250\;nm$ 
spectral region.

Finally, it has been suggested that HAC grains are the carrier of ERE
observed in a variety of astrophysical environments (Duley, Seahra, \&
Williams 1997\markcite{dsw97}). More recently, however, Gordon, Witt,
\& Friedmann (1998)\markcite{gwf98} and Szomoru \& Guhathakurta
(1998)\markcite{sg98} have shown that ERE is also produced by dust in
the diffuse ISM with a quantum efficiency so high as to call into
question its HAC origin. By comparing the absorbed
fraction of the interstellar radiation field, integrated over the
$91.2\!\!-\!\!550\;nm$ range, with the observed ERE intensity over the
same lines of sight, Gordon, Witt, \& Friedmann~(1998)\markcite{gwf98}
conclude and Szomoru \& Guhathakurta confirmed that $(10 \pm 3)\%$ of
all absorbed photons lead to the production of ERE photons.  Since the
ERE carrier is most likely only one of several dust components
contributing to the absorption in the UV/visible wavelength range, the
true quantum efficiency of the ERE process must be larger than 10\%,
and could be as high as 40\% to 50\%.
 
The quantum efficiency of PL in HAC (Silva \etal~1996\markcite{srr96};
Rusli, Robertson, \& Amaratunga 1996\markcite{r96}; Witt, Ryutov, \&
Furton 1997\markcite{wrf97}) is closely correlated with the bandgap,
which in turn is correlated with the H/C ratio. High-efficiency PL
occurs in HAC only if the bandgap is large, of order $3\;eV$, which
implies, under UV illumination, blue-green luminescence instead of red
emission. The absence of blue emission in astronomical sources (Rush
\& Witt 1975\markcite{rw75}; Witt \& Boroson 1990\markcite{wb90})
argues thus against the presence of such high-bandgap HAC and the
large H/C ratios this would imply. Materials such as the HAC sample
discussed in this paper can contribute to the ERE in the correct
wavelength range, but only with a modest quantum efficiency of about
5\%. At best, it would be a partial contributor to the ERE observed in
the diffuse ISM. Other possible sources for the ERE, such as silicon
nanoparticles (Witt, Gordon, \& Furton 1998\markcite{wgf98}; Ledoux
\etal~1998\markcite{l98}) or large PAH molecules (d'Hendecourt
\etal~1986\markcite{h86}) must therefore be considered as main
contributors.  Further, the absence of blue emission in ERE sources
also argues against the presence of HACs with H/C$\;> 0.5$. This
limits at the low end the amount of carbon contained in interstellar
HAC grains because of the existence of the correlation between the H/C
ratio and the band mass-absorption coefficient (see Table~\ref{t5}).
We conclude, therefore, that the amount of carbon locked up in HAC is
nearer $80\;ppM$ than $20\;ppM$ of hydrogen.

\section{Summary}

In summary, we have produced and thoroughly characterized a HAC sample
that is, on the basis of a {\em quantitative} comparison of the IR
absorption spectra of the HAC and of the diffuse ISM, a viable analog
to the true interstellar HAC material.  Both spectra are dominated by
the C-H stretching feature at $\chcm$, with much weaker absorption in
the $2000\!-\!1000\;\icm$ range due to C-C stretching and C-H wagging
modes (in addition to the Si-O band in the interstellar spectrum).
This HAC has a density of $1.5\;g/cm^3$, an electronic band-gap of
$1.9 \pm 0.1 \;eV$, H/C and $sp^3/sp^2$ ratios both near $0.5$, a peak
mass-absorption coefficient in the $\chcm$ band of
$\kappa^{\prime}(2950) = 1.4\EE{3}\;cm^2/g$, and an integrated PL
quantum efficiency of $0.05 \pm 0.02$ with peak emission near
$700\;nm$.  We have also determined via spectroscopic ellipsometry the
complex index of refraction for this HAC in the wavelength range
$300\!-\!1000\;nm$, finding $n$ to be nearly constant near $1.7$ and
$k$ to rise exponentially from near zero in the visible to around
$0.4$ in the near-UV.

We carefully analyzed the profile and strength of the $\chcm$ C-H
stretching absorption band in HAC and the diffuse ISM.  We review the
results of Jacob \& Unger (1996)\markcite{ju96}, showing that the
intrinsic strength of the $\chcm$ feature per C-H bond, and thus the
amount of carbon that needs to be bound in HAC in the diffuse ISM, is
a strong function of the degree of hydrogenation of the material.
HACs with H/C$\;\simeq 0.5$, like the sample described in this work,
are minimally able to provide the observed optical depth in the
interstellar $\chcm$ feature, requiring approximately $80\;ppM$ carbon
relative to hydrogen -- essentially all the available solid-phase
carbon.  HACs with H/C much lower than $0.5$ require increasingly more
than $80\;ppM$ of carbon relative to hydrogen, and thus do not
appear to be viable interstellar HAC analogs.  HACs with H/C much
higher than $0.5$, while able to account for the $\chcm$ feature
optical depth with less carbon, perhaps as little as $20\;ppM$
of hydrogen, also do not appear to be viable analogs because these
materials as a rule exhibit blue-green photoluminescence with an
efficiency on the order of $10\%$ -- such dust emission has never been
observed.  So, we conclude that interstellar HAC must be hydrogenated
to a level of about $0.5\!-\!0.7$ and that this material locks up
most of the available solid-phase carbon.

Finally, we consider other optical roles interstellar HAC is likely to
play in the diffuse ISM.  We conclude that HAC is not the dominant
source of the ERE band, which has now been conclusively detected in
the diffuse ISM, because HAC's PL spectrum and quantum efficiency
differ significantly from ERE.  We also conclude that HAC is not the
sole source of opacity in the visible and near-UV because HACs with $k
>> 0$ in this spectral region require far more carbon than is
available in the solid phase to account for the observed $\chcm$ band
optical depth.  In addition, it is important to note that the
bulk-solid HAC which we have produced does not show any structure in
the spectral vicinity of $220\;nm$.

\acknowledgments

The authors are grateful to Yvonne Pendleton and Doug Whittet for
providing observational data, Margaret Tuma for assistance with
the spectroscopic ellipsometry of our HAC sample, and Thomas Henning 
for a thorough critique of the manuscript.  This work was
supported by NASA grants to the University of Toledo and by
subcontract to Rhode Island College.


\begin{references}
%
\reference{a98}Adamson, A.J., Whittet, D.C.B., Chrysostomou, J.H.,
Hough, J.H., Aitken, D.K., Wright, G.S., Roche, P.F. 1998, \apj,
(submitted)
%
\reference{awd90}Adamson, A.J., Whittet, D.C.B., \& Duley, W.W. 1990,
\mnras, 243, 400
%
\reference{a73}Allen, C.W. 1973, Astrophysical Quantities (London: Athlone)
%
\reference{a91}Alterovitz, S.A., Sauvides, N., Smith, F.W., \&
Woolam, J.A. 1991, in Handbook of Optical Constants II, ed. E.A.
Pallik (New York: Academic), 837
%
\reference{ah88}Angus, J.C. \& Hayman, C.C. 1988, Science, 241, 913
%
\reference{a85}Aspnes, D.E. 1985, in Handbook of Optical Constants of
Solids I, ed. E.D. Palik, Academic Press, 89
%
\reference{aj88}Angus, J.C., \& Jansen F. 1988, J. Vac. Sci. Technol., A6, 1778
%
\reference{b75}Bellamy, L.J. 1975, The Infrared Spectra of Complex
Molecules (New York: Wiley), 3rd edition
%
\reference{b95}Bernstein, M.P., Sandford, S.A., Allamandola, L.J.,
\& Chang, S. 1995, \apj, 454, 327
%
\reference{b85}Rorghesi, A., Bussoletti, E. \& Colangeli 1985, \aap, 142, 225
%
\reference{b87}Bussoletti, E., Colangeli, L. Borghesi, A., \&
Orofino, V. 1987, \aap (supplement), 70, 257
%
\reference{bcc77}Brodsky, M.H.,  Cardona, M., \& Cuomo, J.J.  1977,
\prb, 16, 3556
%
\reference{c96}Cardelli, J.A., Meyer, D.M., Jura, M., \& Savage, B.,
1996, \apj, 467, 334
%
\reference{c98}Chiar, J.E., Pendleton, Y.J., Gaballe, T.R., \&
Tielens, A.G.G.M. 1998, \apj, 507, 281
%
\reference{cg90}Chlewicki, G., \& Greenberg, J.M. 1990, \apj, 365, 230
%
\reference{c93}Colangeli, L., \etal\ 1993, \apj, 418, 435
%
\reference{c95}Colangeli, L., Mennella, V., Palumbo, P., Rotundi, A.,
\& Bussoletti 1995, \aap (supplement), 113, 561
%
\reference{h86}d'Hendecourt, L.B., Leger, A., Olofsson, G. \& Schmidt, W.
1986, \aap, 170, L21
%
\reference{dbk83}Dischler, B., Bubenzer, A., \& Koidl, P. 1983, \apl, 42, 636
%
\reference{d84}Duley, W.W. 1984, \apj, 287, 694
%
\reference{d85}Duley, W.W. 1985, \mnras, 215, 259 
%
\reference{d94}Duley, W.W. 1994, \apjl, 430, L133
%
\reference{d95}Duley, W.W. 1995, in The Diffuse Interstellar Bands,
eds. A.G.G.M. Tielens \& T.P. Snow, (Dordrecht: Kluwer), 359
%
\reference{d98}Duley, W.W., Scott, A.D., Seahra, S., \& Dadswell,
G. 1998, \apjl, 503, L183
%
\reference{dsw97}Duley, W.W.,  Seahra, S., \& Williams, D.A. 1997, ???
%
\reference{gcp88}Gonzalez-Hernandez, J., Chao, B.S., \& Pawlik, D.A. 
1988, \jvst, A7, 2332
%
\reference{gwf98}Gordon, K.D., Witt, A.N., \& Friedmann, B.C. 1998,
\apj, 498, 522
%
\reference{g95}Greenberg, J.M., Li, A., Mendoza-Gomez, C.X.,
Schutte, W.A., Gerakins, P.A., \& De Groot, M. 1995, \apjl, 455, L177
%
\reference{ju96}Jacob, W., \& Unger, M. 1996, \apl, 68, 475
%
\reference{j98}Jaeger, C., Mutschke, H., \& Henning, Th. 1998, \aap, 332, 291
%
\reference{j93}Jenniskens, P. 1993, \aap, 274, 653
%
\reference{jdw90}Jones, A.P.,  Duley, W.W., \& Williams, D.A. 1990, 
\qjras, 31, 567
%
\reference{kjm85}Kaplan, S., Jansen, F., \& Machonkin, M. 1985, 
\apl, 47, 751
%
\reference{l98}Ledoux, G., \etal\ 1998, \aap, 333, L39
%
\reference{lg97}Li, A., \& Greenberg, J.M. 1997, \aap, 323,566
%
\reference{m94}Mathis, J.S. 1994, \apj, 422, 176
%
\reference{m96}Mathis, J.S. 1996, \apj, 472, 643
%
\reference{m89}McFadzean, A.D., Whittet, D.C.B., Longmore, A.J., Bode,
M.F., \& Adamson, A.J. 1989, \mnras, 241, 873
%
\reference{m83}McKenzie, D.R., McPhedran, R.C., Sauvides, N.,
\& Botten, L.C. 1983, Philosophical Magazine B, 48, 341
%
\reference{m98}Mennella, V., Colangeli, L., Bussoletti, E., Palumbo, P.,
\& Rotundi, A. 1998, \apj, 507, L177
%
\reference{od88}Ogmen, M., \& Duley, W.W. 1988, \apj, 334, L117
%
\reference{p94}Pendleton, Y.J., Sandford, S.A., Allamandola, L.J.,
Tielens, A.G.G.M., \& Sellgren, K. 1994, \apj, 437, 683
%
\reference{r86}Robertson, J. 1986, Advances in Physics, 35, 317
%
\reference{rw75}Rush, W.F., \& Witt, A.N. 1975, \aj, 80, 31
%
\reference{r96}Rusli, Robertson, J., \& Amaratunga, G.A.J. 1996, \jap,
80, 2998
%
\reference{s91}Sandford, S.A., Allamandola, L.J., Tielens, A.G.G.M.,
Sellgren, K., Tapia, M., \& Pendleton, Y.J. 1991, \apj, 371, 607
%
\reference{s98}Schnaiter, M., Mutschke, H., Dorschner, J., \&
Henning, Th. 1998, \apj, 498, 486
%
\reference{srr96}Silva, S.R.P., Robertson, J., Rusli, Amaratunga, G.A.J.,
\& Schwan, J. 1996, Phil. Mag. B, 74, 369
%
\reference{s84}Smith, F.W. 1984, J. Appl. Phys., 55, 764
%
\reference{sw95}Snow, T.P., \& Witt, A.N. 1995, Science, 270, 1455
%
\reference{srm76}Soifer, B.T., Russell, R.W., \& Merrill, K.M. 1976, \apjl, 207, L83
%
\reference{sg98}Szomoru \& Guhathakurta 1998, \apj, 494, L93
%
\reference{t73}Tauc, J. 1973, in Amorphous Semiconductors,
ed. M.H. Brodsky (New York:Plenum)
%
\reference{th89}Tamor, M.A., Haire, C.H., Wu, C.H., \& Hass,
K.C. 1989, \apl, 54, 123
%
\reference{tw90}Tamor, M.A., \& Wu, C.H. 1990, J. Appl. Phys., 67, 1007
%
\reference{tw89}Tamor, M.A., Wu, C.H., Carter, R.O. III, \& Lindsay, N.E.
1989, \apl, 55, 1388
%
\reference{tb87}Tsai, H., \& Bogy, D.B. 1987, \jvst, A5, 3287
%
\reference{whk82}Watanabe, I., Hasegawa, S., \& Kurata, Y. 1982, 
Jap. Journal of Appl. Phys., 21, 856
%
\reference{w97}Whittet \etal\ 1997, \apj, 490, 729
%
\reference{w94}Witt, A.N. 1994, in Novel Forms of Carbon II,
eds. C.L. Renschler, D.M. Cox, J.J. Pouch and Y. Achiba,
(Pittsburg:~ Materials Research Society), p.~3
%
\reference{wb90}Witt, A.N., \& Boroson, T.A. 1990, \apj, 355, 182
%
\reference{wgf98}Witt, A.N., Gordon, K.D., \& Furton, D.G. 1998,
\apjl, 501, L111
%
\reference{wrf97}Witt, A.N., Ryutov, D., \& Furton, D.G. 1997, in From
Stardust to Planetesimals, APS Conference Series, eds. Y.J. Pendleton
and A.G.G.M. Tielens, 122
%
\reference{ws88}Witt, A.N., \& Schild, R.E. 1988, \apj, 325, 837
%
\end{references}
\end{document}